\documentclass[submission, PhysCore]{SciPost}

\usepackage{bm}
\usepackage[english]{babel} 

\usepackage{multicol} 
\usepackage{ragged2e} 
\usepackage{quoting} 

\usepackage{microtype} 
\usepackage{textcomp} 
\usepackage[autostyle]{csquotes} 

\usepackage{mathtools} 
\usepackage{amsfonts} 
\usepackage{amssymb} 

\usepackage{dsfont}
\usepackage{physics}
\usepackage{mathrsfs} 

\usepackage{graphicx} 
\usepackage{pdfpages} 
\graphicspath{{./ExterneDateien/Grafiken/},{./ExterneDateien/Dokumente/},{.ExterneDateien/Plots}}

\usepackage{booktabs} 
\usepackage{multirow} 

\usepackage{tikz} 
\usepackage{pgfplots} 
\usepackage{float}
\usepackage{gensymb}

\definecolor{darkred}{RGB}{140, 0, 26} 
\hypersetup{
  linktoc    = all,
  colorlinks = true,
  citecolor  = black, 
  linkcolor  = darkred,
  urlcolor   = blue, 
  filecolor  = blue
}

\usepackage{caption}
\usepackage{subcaption}
\usepackage[capitalise]{cleveref}

\crefname{section}{Sec.}{sections}
\crefname{appendix}{App.}{appendizes}
\usepackage[super]{nth}
\usepackage{color} 

\newif\ifprint
\printfalse

\numberwithin{equation}{section}

\pgfplotsset{compat=1.18}

\begin{document}

\begin{center}
{\Large \textbf{Geometry Dependence of Error Thresholds in Two-Dimensional Toric Codes}}
\end{center}

\begin{center} \textbf{
Daniel Lessing\textsuperscript{1*}, 
Anja Langheld\textsuperscript{1}, 
Calvin Krämer\textsuperscript{1}, 
Jan Alexander Koziol\textsuperscript{2} and 
Kai Phillip Schmidt\textsuperscript{1$\dagger$}}
\end{center}

\begin{center}
\textsuperscript{\textbf{1}} Lehrstuhl für Theoretische Physik V, Friedrich-Alexander-Universität Erlangen-Nürnberg, Staudtstr.~7, 91058 Erlangen, Germany \\
\textsuperscript{\textbf{2}} Faculty of Physics, University of Vienna, Boltzmanngasse 5, 1090 Vienna, Austria
\\
$^*$\href{mailto:daniel.lessing@fau.de}{daniel.lessing@fau.de}, %
$^\dagger$~\href{mailto:kai.phillip.schmidt@fau.de}{kai.phillip.schmidt@fau.de}, %
\end{center}
\section*{Abstract}
{\bf
We investigate how the lattice geometry influences the error thresholds of  two-dimensional toric codes. 
In the presence of bit- or phase-flip errors, the toric code maps onto the two-dimensional random-bond Ising model (RBIM). 
We determine the critical behaviour of the RBIM using replica-exchange Monte Carlo simulations. 
While previous studies have explored thresholds for general lattice geometries under suboptimal decoders such as minimum-weight perfect matching, the rapid development of near-optimal decoders makes resolving the ultimate, maximum-likelihood code capacity relevant. 
We compute these optimal error thresholds on the Nishimori line for the square, honeycomb, triangular, dice, and kagome lattices. 
Owing to lattice duality, the phase-flip threshold on a given lattice is equivalent to the bit-flip threshold on its dual. 
We find that the optimal thresholds of dual-lattice pairs display a characteristic duality-driven splitting around the self-dual square-lattice, mirroring the qualitative behaviour observed for suboptimal decoders.
While the average coordination number has the strongest impact on the error threshold, our results demonstrate that the detailed arrangement of vertices and plaquettes also plays a significant role in determining its precise value.
}

\vspace{0.5cm}
\hrule
\tableofcontents
\vspace{0.5cm}
\hrule
\section{Introduction}
Quantum computers promise to outperform their classical counterparts in a variety of computational tasks \cite{feynman1982simulating, arute2019quantum}. 
However, their practical realization is fundamentally limited by decoherence and unavoidable imperfections in quantum operations \cite{schlosshauer2019quantum}. 
Since the no-cloning theorem prohibits the direct application of copy-based classical error-correction schemes to quantum information, fault-tolerant quantum computation requires fundamentally different strategies. 
Quantum error correction (QEC) overcomes this obstacle by redundantly encoding logical information into entangled many-body states \cite{wootters1982single, kitaev2003fault, terhal2015quantum}, thereby enabling the detection and correction of errors without destroying the encoded quantum information.

An important question in QEC is how the robustness of a code depends on its underlying geometry and dimensionality. 
According to the threshold theorem, logical errors can in principle be suppressed to arbitrarily low levels provided that the physical error rate is below a finite error threshold \cite{aharonov2008fault,knill1997resilient}. 
Determining the factors that govern this threshold is therefore essential for the design of scalable quantum architectures. 
Among the most prominent QEC schemes are topological Calderbank-Shor-Steane codes \cite{Calderbank_1996,Steane_1996}, including the toric code \cite{kitaev2003fault} and color code \cite{Bombin_2006}, which provide a natural framework for fault-tolerant quantum memories and quantum computation in both two and three spatial dimensions. 
Beyond these conventional topological codes, three-dimensional fracton codes have recently emerged as an alternative class of topological stabilizer codes, exhibiting restricted quasiparticle mobility and enhanced protection against thermal and coherent errors \cite{Song2022,Canossa2026}.
Understanding how geometry and dimensionality affect the error thresholds of such topological codes is therefore of broad interest for the development of scalable fault-tolerant quantum technologies.

For the paradigmatic toric code \cite{kitaev2003fault}, the error threshold for the maximum likelihood decoder has been established accurately on the square \cite{kitaev2003fault,dennis2002topological,wang2003confinement} and cubic lattice \cite{Xu2026}, whereas its dependence on more general lattice geometries remains less explored \cite{dequeiroz2009location,ohzeki2009locations}.
Since the lattice geometry directly dictates the connectivity of the physical qubits, it introduces a crucial trade-off between error tolerance and experimental viability.
High-coordination lattices generally offer higher theoretical bit-flip error thresholds, but demand more complex control architectures that increase crosstalk and hardware overhead.
Conversely, lattices with lower connectivity simplify implementation and mitigate parasitic interactions, but have an overall lower threshold.
Thus, understanding the influence of the lattice geometry on the error threshold of bit- and phase-flip errors is both scientifically interesting and technologically relevant. 
Interestingly, also the quantum robustness of the toric code against quantum fluctuations depends strongly on the geometry, i.e.\,, the topological order is more stable on lattices like the triangular and kagome lattices as well as their duals, whenever the dynamics of excitations is geometrically frustrated \cite{schmidt2013persisting, kott2024}.
In addition, it has been shown for the minimum weight perfect matching (MWPM) decoder that the error tolerances against bit- and phase-flip errors depend asymmetrically on the underlying lattice geometry \cite{fujii2012error}.

While suboptimal algorithms like MWPM or union find \cite{Delfosse2021} are widely employed for fast real time decoding, near optimal techniques such as tensor network decoders \cite{bravyi2014efficient} have enabled practical implementations of maximum likelihood decoding. 
Resolving the ultimate maximum likelihood code capacity across general lattice geometries is therefore essential.
It provides a universal baseline to evaluate the performance loss of approximate decoders and reveals the intrinsic topological protection afforded by the lattice itself.

Recent experimental advances have brought topological quantum error correction from a purely theoretical concept to an increasingly realistic hardware platform.
Repeated quantum error correction and logical qubits have recently been demonstrated on superconducting quantum processors \cite{Google2023,Google2024} as well as on trapped-ion quantum computers \cite{Quantinuum2024}. 
At the same time, programmable neutral-atom quantum processors have emerged as a particularly promising architecture for realizing topological stabilizer codes \cite{Bluvstein26} owing to their excellent coherence properties, high-fidelity gate operations \cite{Evered2023}, and the ability to dynamically rearrange atoms into arbitrary two-dimensional geometries \cite{Bluvstein2024,Levine2023}. 
In contrast to fixed-connectivity superconducting circuits \cite{Satzinger2021}, neutral-atom arrays \cite{Bluvstein2022,Bluvstein2024} as well as trapped-ion platforms offer remarkable flexibility in shuttling and engineering lattice structures including periodic boundary conditions \cite{Haghshenas2026}, enabling the direct implementation and comparison of different code geometries within the same experimental platform.

This unique capability makes understanding the influence of lattice geometry on the intrinsic error threshold not only of fundamental theoretical interest but also highly relevant for the optimization of future fault-tolerant quantum hardware.

In this work, we investigate the influence of lattice geometry on the error threshold of the two-dimensional toric code by exploiting its mapping onto the two-dimensional random-bond Ising model (RBIM) \cite{dennis2002topological}. 
Under this mapping, the error threshold for the optimal maximum likelihood decoder is given by the multicritical point at which the RBIM phase boundary intersects with the Nishimori line. 
We determine the phase boundary using large-scale replica-exchange Monte Carlo simulations \cite{dequeiroz2009location, hukushima1996exchange, hasenbusch2008critical,xu2025error}, allowing us to obtain ``maximum-likelihood '' error thresholds for a range of two-dimensional lattice geometries and to quantify the impact of lattice connectivity on the robustness of topological quantum memories.

This article is structured as follows: 
In Section \ref{sec:Toric Code} we define the toric code on various two-dimensional lattice geometries and describe its error correcting properties. 
The mapping of the toric code in the presence of bit- or phase-flip errors to the random bond Ising model is described in Sec.~\ref{sec:RBIM}.
The set up of the replica-exchange Monte Carlo simulations as well as the discussion of the computed error thresholds for the toric code on the square, honeycomb, triangular, dice, and kagome lattice is contained in Sec.~\ref{sec:Results}.
We summarize and conclude our work in Sec.~\ref{sec:Conclusions}.

\section{Toric Code}\label{sec:Toric Code}
\begin{figure}[hbt]
    \centering
    \includegraphics[width=0.5\textwidth]{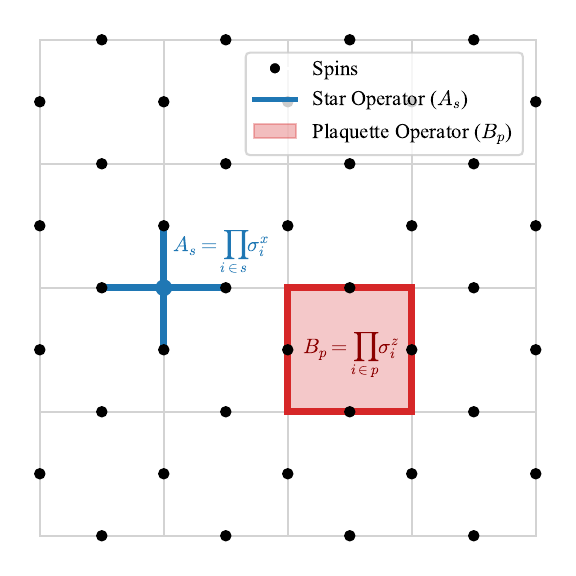}
    \caption{Star ($A_s$) and plaquette ($B_p$) operators on the square lattice.}
    \label{fig:Toric_Code_Architecture}
\end{figure}

The toric code, introduced by Kitaev \cite{kitaev2003fault}, was originally defined on a two-dimensional square lattice with periodic boundary conditions as illustrated in Fig.~\ref{fig:Toric_Code_Architecture}, embedding the system on a torus. 
The physical spin-$1/2$ degrees of freedom reside on the $2N$ links of the lattice, where $N$ is the number of vertices (stars) and plaquettes.

\subsection{Hamiltonian and Ground State}
\begin{figure}[htbp]
    \centering
    \begin{subfigure}[b]{0.45\textwidth}
        \centering
        \includegraphics[]{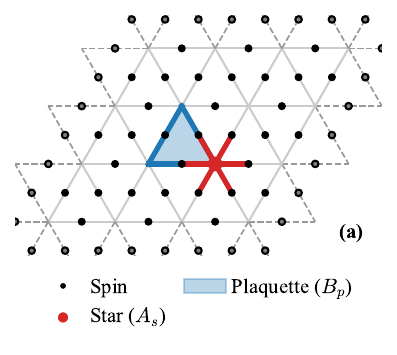}
        \label{fig:Toric Code architexture Tri}
    \end{subfigure}
    \hfill 
    \begin{subfigure}[b]{0.45\textwidth}
        \centering
        \includegraphics[]{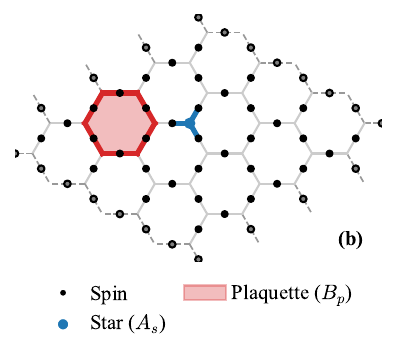}
        \label{fig:Toric Code architecture Hex}
    \end{subfigure}
    \caption{\textbf{(a)} Toric code 6-spin operators $A_s$ and 3-spin operators $B_p$ on the triangular lattice with periodic boundary conditions. \textbf{(b)} Toric code 3-spin operators $A_s$ and 6-spin operators $B_p$ on the honeycomb lattice with periodic boundary conditions.}
    \label{fig:Toric Code Tri and Hex}
\end{figure}

\begin{figure}[htbp]
    \centering
    \begin{subfigure}[b]{0.45\textwidth}
        \centering
        \includegraphics[]{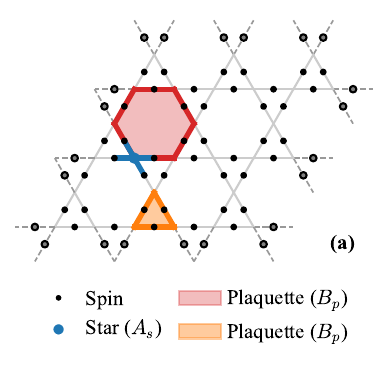}
        \label{fig:toric code kagome lattice}
    \end{subfigure}
    \hfill
    \begin{subfigure}[b]{0.45\textwidth}
        \centering        
        \includegraphics[]{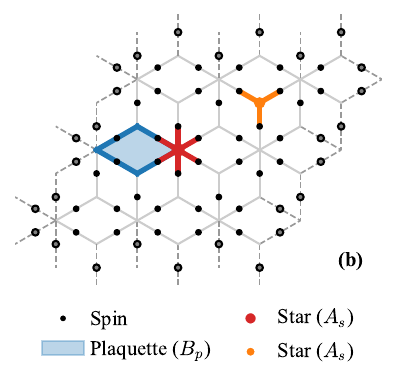} 
        \label{fig:toric code dice lattice}
    \end{subfigure}
    \caption{\textbf{(a)} Star and plaquette operators on a kagome lattice. The lattice features two distinct plaquette types: triangles and hexagons. \textbf{(b)} Star and plaquette operators on a dice lattice. This geometry exhibits star operators of varying sizes.}
    \label{fig:Toric Code Kagome Dice} 
\end{figure}

The toric code can be constructed for an arbitrary two-dimensional lattice geometry on a torus \cite{kitaev2003fault}. 
The Hamiltonian is given by the sum of commuting local stabilizers $A_s$ and $B_p$
\begin{equation}\label{eq:Hamiltonian toric code}
    H_{\rm TC} = -J_s\sum_s A_s -J_p\sum_p B_p\ ,
\end{equation}
with the coupling constants $J_s,J_p > 0$. 
The star operators $A_s$ and plaquette operators $B_p$ are defined as products of Pauli matrices acting on adjacent links

\begin{equation}\label{eq: Operators on the Toric code}
    A_s = \prod_{i \in s} \sigma^x_i \quad \text{and} \quad B_p = \prod_{i \in p} \sigma^z_i\ . 
\end{equation}
The ground-state manifold is defined by the global $+1$ eigenspace of all local stabilizers, forming a degenerate subspace. 
Since local stabilizers cannot distinguish between these degenerate states, non-local Wilson loop operators winding around the non-contractible loops of the torus are required to describe the remaining degrees of freedom
\begin{equation} \label{eq:definition Wilson loop operators}
    W_\gamma^z = \prod_{i \in \gamma}\sigma_i^z \quad \text{and} \quad W_{\tilde{\gamma}}^x = \prod_{i \in \tilde{\gamma}}\sigma_i^x\ , 
\end{equation}
where $\gamma$ and $\tilde{\gamma}$ denote loops on the direct and dual lattice, respectively. 
The topology of the torus yields two independent non-contractible loops.
Generally these loops share an even number of spins with any star or plaquette, and therefore commute with the Hamiltonian, $[H_\mathrm{TC}, W_{\gamma,\tilde{\gamma}}^{z,x}] = 0$. 
Consequently, applying a Wilson loop leaves the system's energy invariant but maps one ground state to another. 
Choosing one set of Wilson loops either on the direct or dual lattice as logical observables, $W_{\gamma_1}^z \equiv Z_1$ and $W_{\gamma_2}^z \equiv Z_2$, defines a complete, orthogonal basis $\{|\lambda_1, \lambda_2\rangle\}$ with $\lambda_{1,2} \in \{+1, -1\}$ for the logical subspace~\cite{kitaev2003fault, KaiTOPO}. 
This framework will be applied across the multiple geometries investigated in this work, specifically the square, honeycomb, triangular, kagome, and dice lattices shown in \cref{fig:Toric_Code_Architecture}, \cref{fig:Toric Code Tri and Hex}, and \cref{fig:Toric Code Kagome Dice}.

\begin{figure}[htbp]
    \centering
    \begin{subfigure}[b]{0.45\textwidth}
        \centering
        \includegraphics[width=\textwidth]{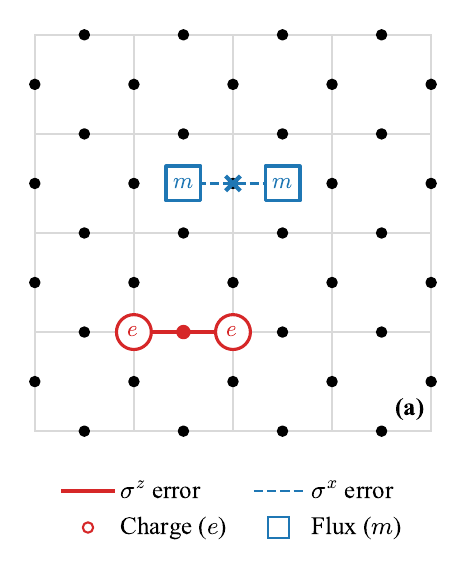}
        \label{fig:adjacent_pair}
    \end{subfigure}
    \hfill 
    \begin{subfigure}[b]{0.45\textwidth}
        \centering
        \includegraphics[width=\textwidth]{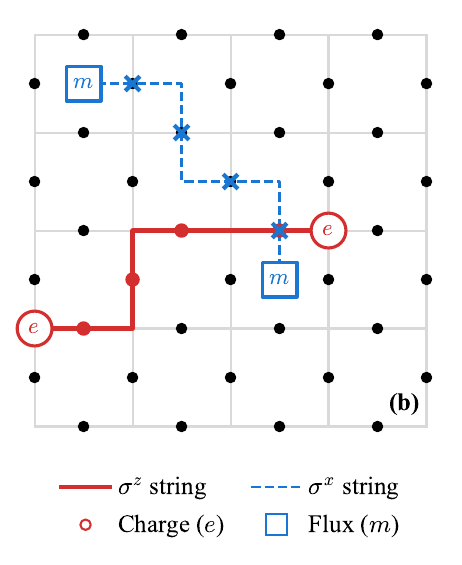}
        \label{fig:separated_excitations}
    \end{subfigure}
    \caption{Movement of excitations in the toric code. \textbf{(a)} Local Pauli operations ($\sigma^z$ or $\sigma^x$) create pairs of charges ($e$) or fluxes ($m$) on neighbouring stars/plaquettes. \textbf{(b)} Applying further operators extends the error string, and moves the quasiparticles to the endpoints.}
    \label{fig:excitations_combined}
\end{figure}

\subsection{Error Correcting Properties}\label{susec:Error Correcting Properties}
In its ground state $\ket{0}$, the toric code occupies a minimum energy configuration where all $A_s$ and $B_p$ stabilizers yield an eigenvalue of $+1$. 
Local Pauli perturbations, acting on a single spin, flip two adjacent stabilizer eigenvalues to $-1$, creating a pair of localized quasiparticle excitations. 
A $\sigma^x$ operation ($X$-type error) induces two $B_p$ magnetic fluxes ($m$), while a $\sigma^z$ operation ($Z$-type error) creates two $A_s$ electric charges ($e$) (see \cref{fig:excitations_combined}).

When a chain of such operations acts on a sequence of neighbouring spins, any stabilizer covering an even number of modified links continues to return an eigenvalue of $+1$. Consequently, the resulting excitations remain localized exclusively at the endpoints of the error chain, leaving the exact error path unknown (see \cref{fig:separated_excitations}).

Because the local stabilizer operators mutually commute, the error syndrome can be extracted non-destructively without altering the stored state, allowing a decoder to annihilate these local excitations before they combine into larger loops.

Since the logical information is encoded in the global topology of the system, defined by the non-local Wilson loop operators in \cref{eq:definition Wilson loop operators}, a logical error only occurs when the error chain combined with the applied correction forms a closed, non-contractible loop around the torus. 
In contrast to open error strings, such a global loop possesses no boundaries and commutes with all local stabilizers, leaving the system in a $+1$ eigenvalue configuration. 
This creates an undetectable logical error that maps the ground state onto a different one. 
Assuming an independent physical error probability $p$ for each qubit, the error threshold $p_\mathrm{th}$ marks the critical limit below which the logical failure rate $P$ is exponentially suppressed with increasing linear system size $L$. 
This threshold will be evaluated in the subsequent sections by mapping the error-correction problem onto a statistical mechanics model.

\section{Toric Code Mapping to the 2D RBIM}\label{sec:RBIM}
It has been well established that the error threshold of different toric code geometries can be found by mapping the quantum error-correction process onto the two-dimensional random bond Ising model (RBIM) \cite{dennis2002topological}. 
This mapping relates the failure rate of a maximum-likelihood decoder to the thermodynamic phase transitions of the RBIM.

\subsection{Two-Dimensional Random Bond Ising Model}
\label{sec: 2D RBIM}
The RBIM is defined by the Hamiltonian \cite{nishimori2001statistical}
\begin{equation}\label{eq:Hamiltonian RBIM}
    H = -\sum_{\langle i, j \rangle} J_{ij}\sigma_i\sigma_j\ ,
\end{equation}
where the sum $\langle i, j\rangle$ runs over the nearest neighbours of the underlying lattice. 
This work focuses on the $\pm J$ model, where the couplings $J_{ij}$ take discrete values of $-J$ or $+J$ according to a bimodal probability distribution, with probabilities $p$ and $1-p$, respectively. 
Following Ref.~\cite{nishimori2001statistical}, the symmetry of this disordered system can be investigated by a local gauge transformation of both spins and interactions at all lattice sites
\begin{equation}
\label{eq:gauge symmetry of 2D RBIM}
    \sigma_i \rightarrow \sigma_i\tau_i\ , \qquad J_{ij} \rightarrow J_{ij}\tau_i\tau_j\ ,
\end{equation}
where $\tau_i \in \{+1, -1\}$ at a site $i$. 
Under this transformation, the product transforms as $J_{ij}\sigma_i\sigma_j \rightarrow J_{ij}\tau_i\tau_j \cdot \sigma_i\tau_i \cdot \sigma_j\tau_j = J_{ij}\sigma_i\sigma_j$, proving that the Hamiltonian remains gauge invariant \cite{nishimori2001statistical}. 
For the configurations considered in this work, the RBIM is evaluated on two-dimensional lattices with periodic boundary conditions.

\subsection{The Statistical Mechanics Mapping}
\begin{figure}[htbp]
    \centering
    \includegraphics[]{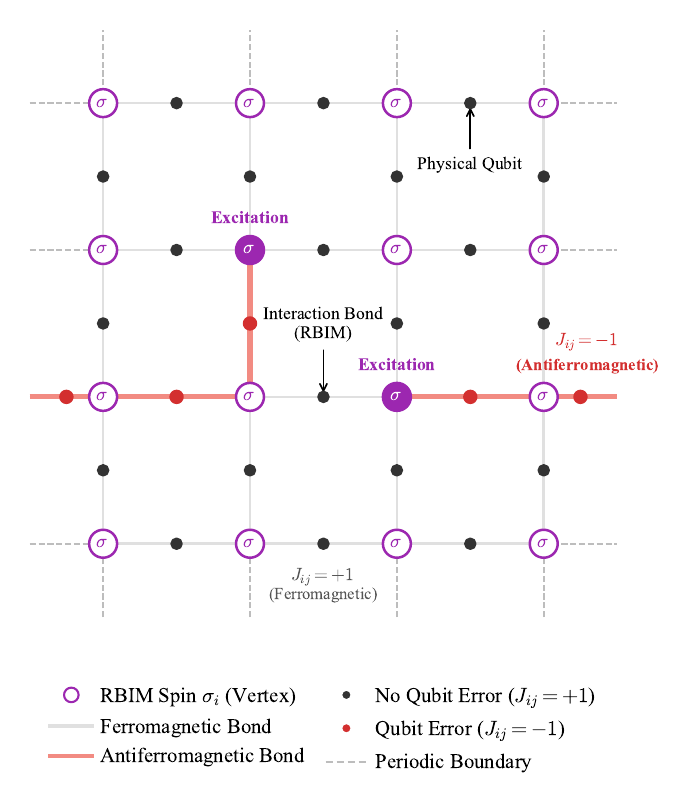}
    \caption{Geometric mapping of the 2D RBIM to the toric code stabilizers for $X$ errors. Qubits sit on the bonds connecting nearest-neighbor Ising spins $\sigma$.}
    \label{fig:RBIM_mapping_TC} 
\end{figure}

For the correction of bit-flip ($X$) errors, the spins of the RBIM reside on the vertices of the dual lattice of the toric code, ensuring that each Ising bond crosses exactly one lattice link (physical qubit). 
In contrast, for phase-flip ($Z$) errors, RBIM spins reside on the vertices of the primal lattice. Consequently, a qubit error corresponds to an antiferromagnetic coupling ($J_{ij} = -J$), while error-free links are mapped to ferromagnetic bonds ($J_{ij} = +J$), as illustrated in \cref{fig:RBIM_mapping_TC}.

An error chain $E$ represents the set of all qubit errors. 
Its boundary defines the error syndrome $S$, which corresponds to the set of frustrated plaquettes in the RBIM. 
Under perfect syndrome measurements, any valid error chain can be decomposed as $E = S' + C$, where $S'$ is an arbitrary reference path matching the boundary $S$, and $C$ is a closed loop \cite{dennis2002topological}. 
Here, $+$ denotes the disjoint union.

\begin{figure}[htbp]
    \centering
    \includegraphics[width=0.45\textwidth]{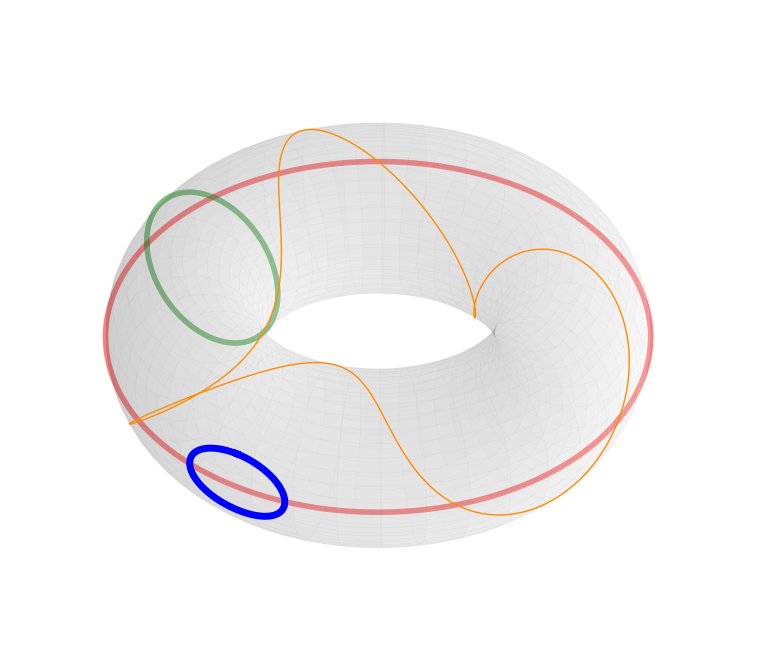}
    \caption{Homology classes on the toric code, representing trivial ($I$) loops in blue and non-trivial ($X, Z, XZ$) loops in red, green, and orange.}
    \label{fig:homology_classes_toric_code}
\end{figure}

Chains belong to the same homology class if they can be transformed into one another via local stabilizer operations. 
On a torus, this classification yields a trivial class $I$ (no logical operation) and three non-trivial classes $L \in \{X, Z, XZ\}$, corresponding to the non-contractible Wilson loop operators (see \cref{fig:homology_classes_toric_code}). 
Error correction succeeds only if the applied correction chain and the actual error chain $E$ share the same homology class, meaning that their combination forms a homologically trivial loop.

For a static error configuration, the conditional probability that a measured syndrome $S$ was caused by a chain belonging to a specific homology class $h$ is given by \cite{dennis2002topological}
\begin{equation}\label{eq:probability_homology_class}
    \mathrm{prob}(h|S) = \frac{\sum_{C' \in h} \mathrm{prob}(S'+C')}{\sum_{C'} \mathrm{prob}(S'+C')}\ .
\end{equation}
By defining the total logical failure probability $P_{\mathrm{logical}}$ as the sum over all non-trivial classes $L$, the decoding problem can be written as a ratio of partition functions
\begin{equation}\label{eq:probability logical error}
    P_{\mathrm{logical}} = \sum_E \mathrm{prob}(E) \frac{\sum_{D \in L} \mathrm{prob}(E+D)}{\sum_{D} \mathrm{prob}(E+D)}\ .
\end{equation}
The summation over the closed chains $D$ maps to the partition function $Z_h(E)$ of the RBIM under a given disorder configuration fixed by the error chain $E$ \cite{dennis2002topological}
\begin{equation}\label{eq: RBIM mapping relation to partition function}
    \sum_{D\in h} \mathrm{prob}(E+D) \propto Z_h(E) = \sum_{\{\sigma\} \in h}e^{-\beta H(\{\sigma\}|E)}\ ,
\end{equation}
where $H(\{\sigma\}|E) = -\sum_{\langle i, j \rangle} J_{ij}\sigma_i\sigma_j$ with $J_{ij} = -1$ on links containing an error.
Substituting \eqref{eq: RBIM mapping relation to partition function} into \eqref{eq:probability logical error} yields
\begin{equation}\label{eq:logical error RBIM complicated}
    P_{\mathrm{logical}} = \sum_E\mathrm{prob}(E) \frac{\sum_{\{ \sigma \} \in L}e^{-\beta H(\{ \sigma \}|E)}} {\sum_{\{ \sigma \}} e^{-\beta H(\{ \sigma \}|E)}} = \sum_L \langle e^{-F_{\mathrm{DW}_L}(E)} \rangle_E\ ,
\end{equation}
where $F_{\mathrm{DW}_L}(E) \equiv -\ln [Z_L(E)/Z_{\mathrm{total}}(E)]$ defines the domain wall (DW) free energy cost to create a percolating error chain with a non-trivial homology class $L$ \cite{dennis2002topological}.

\begin{figure}[htbp]
    \centering
    \includegraphics[]{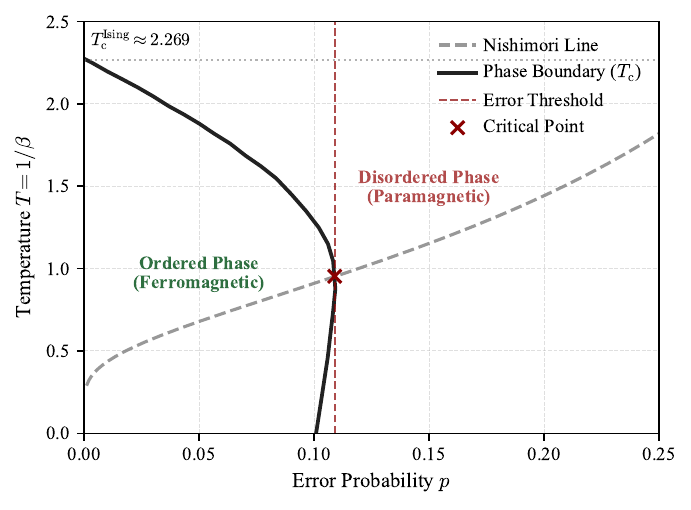}
    \caption{Qualitative phase diagram of the $\pm J$ RBIM on the square lattice. The threshold $p_{\mathrm{th}}$ is located at the intersection of the phase boundary and the Nishimori line.}
    \label{fig:RBIM_phase_diagram}
\end{figure}

To match the maximum-likelihood decoding condition, the independent physical error probability $p$ must be mapped to the Boltzmann weights of the spin model. 
This fixes the system onto the Nishimori line \cite{nishimori1980exact}
\begin{equation}\label{eq:nishimori_condition}
    e^{-2\beta J} = \frac{p}{1-p} \quad \Leftrightarrow \quad \beta = \frac{1}{2J} \ln\left(\frac{1-p}{p}\right)\ .
\end{equation}
The error threshold $p_{\mathrm{th}}$ thus corresponds to the exact point where the phase boundary of the RBIM intersects the Nishimori line (see \cref{fig:RBIM_phase_diagram}).
The local gauge symmetry $\sigma_i \rightarrow \sigma_i\tau_i$, $J_{ij} \rightarrow J_{ij}\tau_i\tau_j$ defined in \cref{eq:gauge symmetry of 2D RBIM} reflects the local stabilizer invariance of the toric code. 
Applying a stabilizer operator shifts the local configurations but leaves the global free energy landscape invariant.

Because homology classes and the creation of a domain wall are topological features, the formulation of the statistical mechanics mapping is independent of the underlying lattice architecture \cite{kitaev2003fault, dennis2002topological}. 
The relation between error strings and the partition function \eqref{eq: RBIM mapping relation to partition function} remains valid. 
Generally, the summation index $\langle i,j \rangle$ in $H(\{\sigma\}|E)$ tracks the nearest neighbours of the chosen coordinate lattice. 
Since the scaling of $F_{\mathrm{DW}_L}$ depends on the local lattice connectivity, the numerical value of the threshold $p_{\mathrm{th}}$ shifts fundamentally across different geometries.

For the RBIM on mutually dual lattices, Ref.\,\cite{takeda2005exact} conjectured that the critical error thresholds $p_{\mathrm{c}}$ on the primary lattice and $p_{\mathrm{c}}^*$ on the dual lattice intersecting the Nishimori line satisfy the binary entropy duality constraint
\begin{equation}\label{eq:nishimori_duality}
    H_2(p_{\mathrm{c}}) + H_2(p_{\mathrm{c}}^*) = 1\ ,
\end{equation}
where $H_2(p) = -p \log_2(p) - (1-p) \log_2(1-p)$. 
For the self-dual square lattice ($p_{\mathrm{c}} = p_{\mathrm{c}}^*$), \cref{eq:nishimori_duality} yields $H_2(p_{\mathrm{c}}) = 1/2$, giving $p_{\mathrm{c}} \approx 0.1100$ \cite{ohzeki2008duality} (subsequently refined to $0.109187$ \cite{ohzeki2009locations}, matching high-precision Monte Carlo integration \cite{hasenbusch2008critical, honecker2001universality}).

In the context of topological quantum codes, where bit-flip ($X$) and phase-flip ($Z$) errors reside on mutually dual lattices, \cref{eq:nishimori_duality} provides an analytical benchmark to evaluate the numerical consistency of error thresholds across dual geometries.

\section{Numerical Error Threshold Estimations}
\label{sec:Results}
In this section we locate the error threshold $p_{\mathrm{c}}$ of the toric code across various two-dimensional lattice geometries.
We do this by calculating the phase diagrams of the $\pm J$ RBIM via finite-size scaling of data generated via Monte Carlo simulations.
We will first briefly describe the method and benchmark it using the well known threshold of the toric code on the square lattice.
Then we will present and discuss the thresholds of various geometries.

\subsection{Methods and Benchmarking}
To estimate the error thresholds for a given lattice geometry, we employ replica exchange Monte Carlo also known as parallel tempering \cite{hukushima1996exchange, earl2005parallel}. 
For a given error rate, systems at multiple temperatures are simulated while proposing an exchange of the spin configurations of neighbouring temperatures. 
To do this efficiently, the temperatures are distributed geometrically according to a fixed ratio in the inverse temperature space \cite{kone2005selection}.

In this work, linear system sizes up to $L = 48$ (corresponding to $2304$ physical spins on the square lattice) were simulated using $N \approx \frac{3}{4} \cdot L$ temperatures per lattice size. 
Ensemble averaging was conducted for up to $51600$ disorder samples for smaller system sizes and $21600$ disorder samples for the largest system size $L = 48$.

During the simulation we determine moments of the magnetisation to calculate Binder cumulants \cite{binder1981finite}.
According to the finite-size scaling hypothesis \cite{Fisher1972, binder1981finite}, Binder cumulants are defined such that they are independent of their finite system size at the critical point.
We first identify several critical points along the phase boundary near the Nishimori line via Binder cumulant intersections.
We then extrapolate these boundary points to calculate their exact intersection with the Nishimori line, where a dedicated Binder cumulant analysis is performed to pinpoint the threshold.
Additionally, we employ the data-collapse method at the determined temperature of the Nishimori point, which fits the scaling form to the cumulant curves of different system sizes at the same time to get an estimate for both the critical point and the critical exponent $\nu$ of the correlation length.

The detailed phase diagrams for each lattice geometry are presented in more detail in Appendix~\ref{sec:Appendix}.

\begin{figure}[htbp]
    \centering
    \begin{subfigure}[b]{0.75\textwidth}
        \centering
        \includegraphics[]{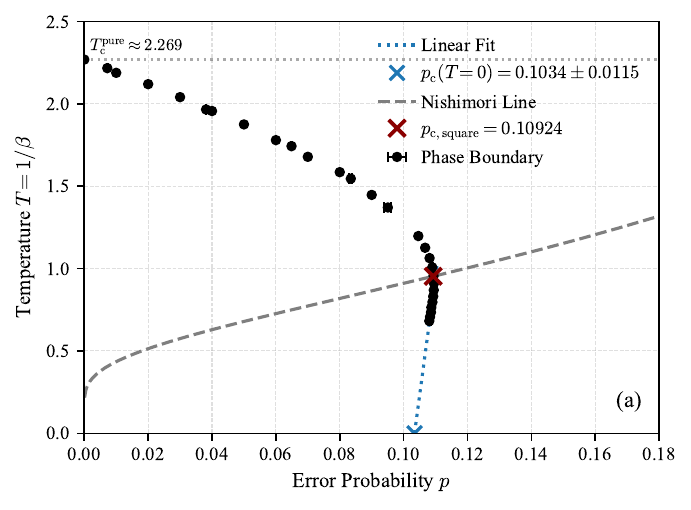}
        \label{fig:phase_line_square_sub}
    \end{subfigure}
    \vspace{0.8em}
    \begin{subfigure}[b]{0.75\textwidth}
        \centering
        \includegraphics[]{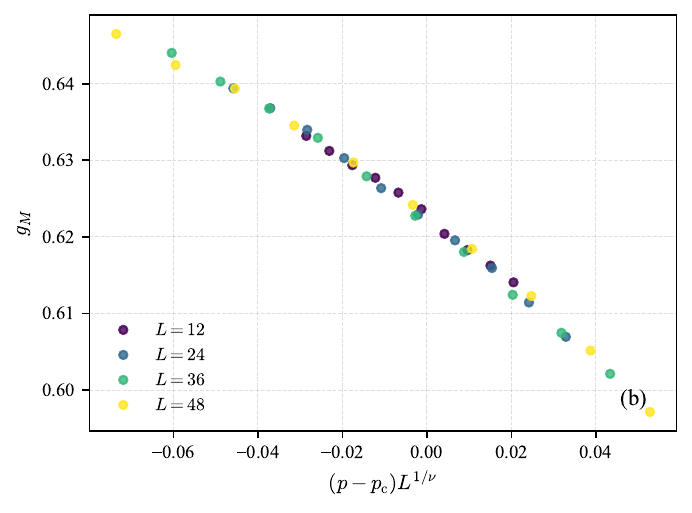}
        \label{fig:data_collapse_square_sub}
    \end{subfigure}
    \caption{Numerical analysis of the $\pm J$ RBIM on the square lattice. \textbf{(a)} Phase boundary of the RBIM determined via replica exchange Monte Carlo. The intersection of the phase boundary and the Nishimori line is found at $T_{\mathrm{c}} = 0.9532$ and \mbox{$p_{\mathrm{c}} = 0.1093 \pm 0.0006$} determined via cubic interpolation. \textbf{(b)} Data collapse of the Binder cumulant at $T_{\mathrm{c}} = 0.9532$, scaling the data onto a single curve yielding a threshold of $p_{\mathrm{c}} = 0.10924 \pm 0.00027$ and critical exponent $\nu = 1.47 \pm 0.17$ using Jackknife error estimation.}
    \label{fig:combined_square_lattice_results}
\end{figure}

To benchmark the procedure, we first apply it to the standard square lattice, where high-precision numerical and analytical results are well-established \cite{nishimori1980exact, nishimori1981internal, nishimori2001statistical, ohzeki2008duality, hasenbusch2008critical, honecker2001universality, dequeiroz2009location, queiroz2006multicritical, ohzeki2009locations, ohzeki2015high, merz2002twodimensional, takeda2005exact}. 
As illustrated in \cref{fig:phase_line_square_sub,fig:data_collapse_square_sub}, the combination of raw cumulant intersections along the phase transition line with a subsequent data collapse robustly yields consistent threshold values $p_{\mathrm{c}}$.

The critical parameters extracted for the square lattice exhibit a good agreement with the established literature. 
The cubic interpolation yields $p_{\mathrm{c}} = 0.1093 \pm 0.0006$ (see Fig.~\ref{fig:combined_square_lattice_results}), aligning with high-precision literature values ($p_{\mathrm{c}} \approx 0.10919$) \cite{hasenbusch2008critical, honecker2001universality} and the analytic conjecture $p_{\mathrm{c}} \approx 0.109187$ \cite{ohzeki2009locations}.
For $T < T_{\mathrm{c}}$, the boundary exhibits re-entrant behavior, extrapolating to a zero-temperature threshold of \mbox{$p_{\mathrm{c}}(T=0)=~0.1034 \pm 0.0115$}, consistent with Ref.~\cite{ohzeki2015high, fujii2012error} ($p_{\mathrm{c}}(T=0) \approx 0.1031$). 
This corresponds to the error threshold for the MWPM decoder investigated in \cite{fujii2012error}. 
In the pure Ising limit ($p=0$), the critical coupling $\beta_{\mathrm{pure}} = 0.44071 \pm 0.00017$ matches the exact Onsager solution ($\beta_{\mathrm{exact}} \approx 0.44069$).

Refining these multicritical parameters via data collapse then yields a threshold of $p_{\mathrm{c}}=0.10924 \pm 0.00027$ (see Fig.~\ref{fig:combined_square_lattice_results}). 
We find that the global optimization of the data collapse provides the most robust parameter estimation. 
By scaling the entire dataset simultaneously, it averages out statistical fluctuations and local fitting errors that degrade conventional intersection analyses. 
Consequently, all final error thresholds reported for the remaining geometries in this work will be strictly oriented toward the values obtained from the data collapse.

\subsection{Error Thresholds}
Using a finite-size scaling analysis, we extract the critical error thresholds $p_\mathrm{c}$ for all investigated geometries (\cref{tab:rbim_results}).
\begin{table}[htbt]
    \centering
    \caption{Numerical results for the RBIM at the Nishimori point obtained via data collapse. $\sum H_2(p_\mathrm{c})$ denotes the sum of the binary entropies for dual lattice pairs.}
    \label{tab:rbim_results}
    \begin{tabular}{lccc}
        \toprule
        \textbf{Lattice} & \textbf{Avg. Coordination} $c$ & \textbf{Error Threshold} $p_\mathrm{c}$ & $\sum H_2(p_\mathrm{c}) \approx 1$ \\
        \midrule
        Square      & 4 & $0.10924 \pm 0.00027$ & 0.99524\\
        \addlinespace
        Honeycomb   & 3 & $0.0654 \pm 0.0007$ & \multirow{2}{*}{0.9907}  \\
        Triangular & 6 & $0.1633 \pm 0.0003$ &  \\
        \addlinespace
        Kagome      & 4 & $0.1000 \pm 0.0005$ & \multirow{2}{*}{0.9911}  \\
        Dice        & 4 & $0.1175 \pm 0.0006$ &  \\
        \bottomrule
    \end{tabular}
\end{table}
Comparing the obtained thresholds reveals the duality-driven splitting centered around the self-dual square lattice ($p_\mathrm{c,sq} \approx 0.1092$). 
Dual-lattice pairs group around this benchmark, deviating in opposite directions according to their connectivity. 
The highly connected triangular lattice ($c=6$) exhibits the highest threshold ($p_\mathrm{c,tri} \approx 0.1633$), whereas its dual honeycomb lattice ($c=3$) yields the lowest \mbox{($p_\mathrm{c,hex} \approx 0.0654$)}. 
These results are slightly lower than previous estimations \cite{dequeiroz2009location, ohzeki2009locations}, displaying characteristic offsets of $\Delta p \approx +0.054$ and $-0.044$ relative to the square-lattice benchmark. 
Overall, the thresholds reported by de Queiroz \cite{dequeiroz2009location} are slightly higher across lattices, which can likely be attributed to methodological differences and finite-size effects.

\begin{figure}[htbt]
    \centering
    \includegraphics[]{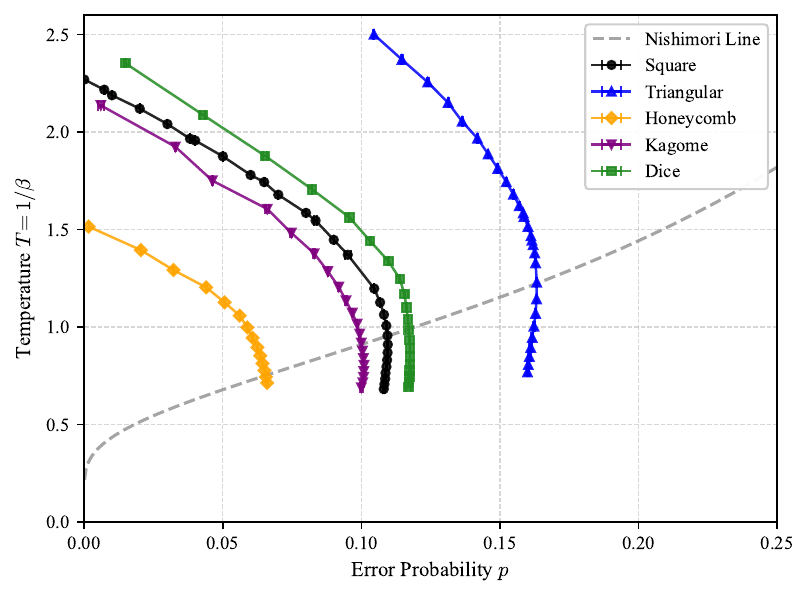}
    \caption{Phase diagrams of all lattice geometries with their Nishimori line intersections. Phase boundaries are obtained via Binder cumulant intersections. The intersection with the Nishimori line has been further refined by a data collapse.}
    \label{fig: phase diagram of all lattice geometries}
\end{figure}

Despite both the dice and kagome lattices having an average coordination number of $c=4$, the dice lattice is slightly more robust ($p_\mathrm{c,dice} \approx 0.1175$) than the kagome lattice ($p_\mathrm{c,kag} \approx 0.1000$), splitting by $\Delta p \approx +0.008$ and $-0.009$ around the square lattice value. 
This duality-driven behavior directly reflects the generalized duality relation $\sum H_2(p_\mathrm{c}) \approx 1$ \cite{ohzeki2008duality, takeda2005exact}, which our numerical estimates satisfy within a $1\%$ margin (see \cref{tab:rbim_results}). 
The complete phase diagrams illustrating this central splitting across all studied geometries are shown in \cref{fig: phase diagram of all lattice geometries}. 

Physically, while the average coordination number $c$ has the main influence on the threshold, the form factor of the unit cell dictates the offset relative to the square lattice threshold. 
The energy cost to create a domain wall, which corresponds to the weight of a logical error chain, scales with vertex connectivity, penalizing error propagation through high connectivity vertices in the triangular and dice geometries. 
In contrast, large plaquettes, such as the six-bond hexagons in the honeycomb and kagome lattices, accommodate more error configurations without inducing frustration, acting as paths of least resistance. 
Because a high-connectivity vertex in a primal lattice directly maps to a large plaquette in its dual, this structural interplay governs the precise threshold splitting around the self-dual square lattice.

\begin{figure}
    \centering
    \includegraphics[]{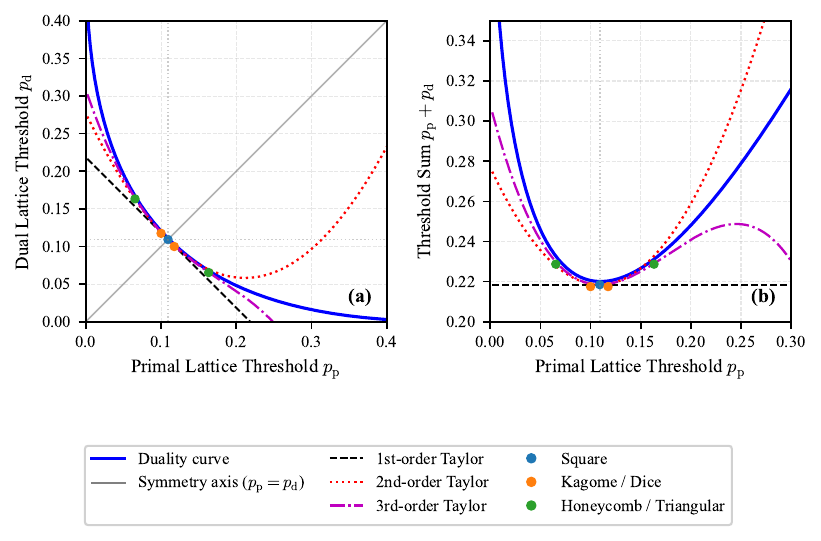}
    \caption{Duality relation and analytical Taylor expansion compared with the numerical thresholds determined via Monte Carlo simulations. Acquired data follows the duality curve within a \(1\%\) margin. \textbf{(a)} Threshold splitting around the self-dual square lattice. 
    \textbf{(b)} Threshold sum as a function of the varying primary lattice threshold.}
    \label{fig:duality_splitting}
\end{figure}

Following the conjecture by \cite{ohzeki2008duality}, the sum of the error thresholds of the primal and dual lattice has a minimum at the self-dual point (see Fig.~\ref{fig:duality_splitting}).
Therefore, any deviation of the error thresholds of this self-dual point leads to an increase in the threshold sum.
This behaviour follows directly from the conjecture \cite{ohzeki2008duality} and can be seen in the numerical solution of Eq.~\eqref{eq:nishimori_duality}.
Furthermore, this can be understood analytically using a Taylor expansion around the self-dual point.
A derivation of the Taylor expansion can be found in Appendix~\ref{app:Taylorexpansion}.
In practice, the threshold sum for non-self dual lattices is in a good approximation equal to the self-dual point for small deviations of the primary lattice threshold.
Therefore, to maximize the threshold sum, a larger deviation from the self-dual point can be beneficial if the smaller threshold on the dual lattice can be accounted for (see Fig.~\ref{fig:duality_splitting}).

\section{Conclusions}\label{sec:Conclusions}

In this work, we have investigated the influence of lattice geometry on the error thresholds of the two-dimensional toric code. By mapping bit- and phase-flip errors onto the two-dimensional random-bond Ising model and employing replica-exchange Monte Carlo simulations, we extended previous zero-temperature studies ($T=0$), which focus solely on the MWPM decoder \cite{fujii2012error}, by mapping the full finite-temperature phase diagrams across general lattice geometries. Locating the multicritical points on the Nishimori line allowed us to resolve the optimal, maximum-likelihood error thresholds for the square, honeycomb, triangular, dice, and kagome lattices. 

Our results demonstrate that the optimal thresholds of dual-lattice pairs display a characteristic duality-driven splitting centred about the self-dual square-lattice benchmark, establishing this mechanism as a fundamental feature of the code capacity. Furthermore, while the average coordination number sets the primary scale for the error threshold, the detailed arrangement of high-connectivity vertices and large plaquettes induces significant shifts, highlighting the decisive role of local microstructures beyond simple connectivity.
Interestingly, the influence of lattice geometry extends beyond error thresholds. Previous studies have shown that the quantum robustness of the toric code against quantum fluctuations, e.g., due to the presence of a magnetic field  also depends strongly on the underlying geometry, with topological order being significantly more stable on geometrically frustrated lattices in field directions, where the quantum dynamics of elementary excitations (errors) is frustrated \cite{schmidt2013persisting,kott2024}. Most extremely this is the case for the toric code on the kagome or dice lattice where the topological order can persist for arbitrary field strength in certain field directions \cite{schmidt2013persisting}.

By establishing lattice geometry as an explicit design parameter for optimizing maximum-likelihood capacity, our findings provide guidance for designing robust, scalable topological quantum memories. Important directions for future work include extending this finite-temperature thermodynamic mapping to three-dimensional toric codes, as well as investigating geometric capacity limits in other topological stabilizer families, such as color codes and fracton architectures.

\section*{Acknowledgements}
We thank Maximilian Vieweg for many fruitful discussions.

\paragraph{Author contributions}
DL: Conceptualization, Data curation, Formal analysis, Investigation, Methodology, Software, Visualization, Writing -- original draft, Writing -- review \& editing.
AL: Conceptualization, Investigation, Methodology, Supervision, Writing -- review \& editing.
CK: Conceptualization, Investigation, Methodology, Supervision, Writing -- review \& editing.
JAK: Conceptualization, Investigation, Methodology, Supervision, Writing -- review \& editing.
KPS: Conceptualization, Funding acquisition, Methodology, Resources, Supervision, Writing -- review \& editing.\footnote{Following the taxonomy \href{https://credit.niso.org}{CRediT} to categorize the contributions of the authors.}

\paragraph{Funding information}
We gratefully acknowledge the scientific support and HPC resources provided by the Erlangen National High Performance Computing Center (NHR@FAU) of the Friedrich-Alexander-Universität Erlangen-Nürnberg (FAU).
The hardware of NHR@FAU is funded by the German Research Foundation (DFG).
The authors gratefully acknowledge the support by the Munich Quantum Valley, which is supported by the Bavarian state government with funds from the Hightech Agenda Bayern Plus.
JAK is funded by the Austrian Science Fund (FWF) [10.55776/COE1, 10.55776/F101200] and the European Union (NextGenerationEU).

\appendix 

\section{Finite-size Scaling and Phase Diagrams}\label{sec:Appendix}
\begin{figure}[htbp]
    \centering
    \begin{subfigure}[b]{0.48\textwidth}
        \centering
        \includegraphics[]{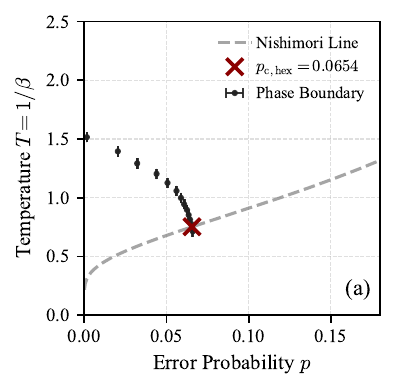}
        \label{fig: Phaseline of the hexagonal lattice}
    \end{subfigure}
    \hfill
    \begin{subfigure}[b]{0.48\textwidth}
        \centering
        \includegraphics[]{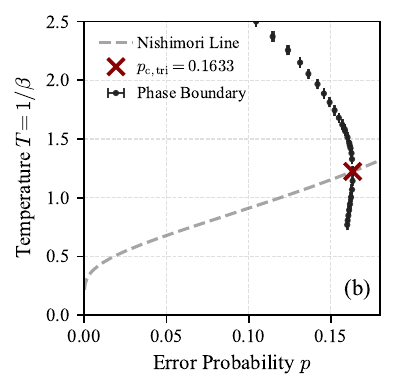}

        \label{fig: Phase line Triangular lattice}
    \end{subfigure}
    \caption{Phase diagrams for the dual pair of \textbf{(a)} honeycomb and \textbf{(b)} triangular lattices. Intersections with the Nishimori line are obtained via cubic interpolation. Numerical parameters are compiled in \cref{tab:appendix_detailed_results}.}
    \label{fig:Phase lines honeycomb triangular}
\end{figure}

\begin{figure}[htbp]
    \centering
    \begin{subfigure}[b]{0.48\textwidth}
        \centering
        \includegraphics[]{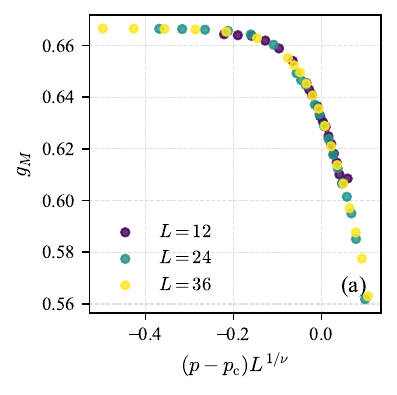}
        \label{fig:data collapse honeycomb lattice}
    \end{subfigure}
    \hfill 
    \begin{subfigure}[b]{0.48\textwidth}
        \centering
        \includegraphics[]{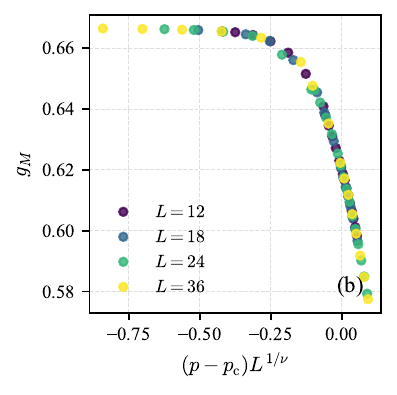}
    \end{subfigure}
    \caption{Finite-size scaling data collapse of the Binder cumulant at the Nishimori point for \textbf{(a)} the honeycomb and \textbf{(b)} the triangular lattice. See \cref{tab:appendix_detailed_results} for the extracted scaling exponents.}
    \label{fig:Data collapse of hexagonal and triangular lattice}
\end{figure}

\begin{figure}[htbp]
    \centering
    \begin{subfigure}[b]{0.48\textwidth}
        \centering
        \includegraphics[]{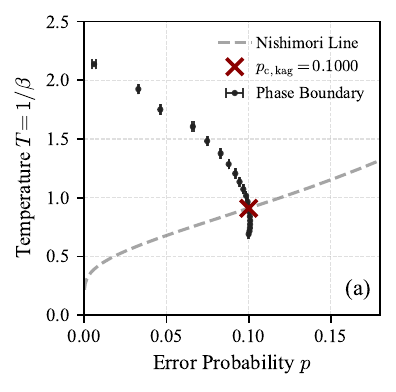}
        \label{fig: phase line for Kagome}
    \end{subfigure}
    \hfill
    \begin{subfigure}[b]{0.48\textwidth}
        \centering
        \includegraphics[]{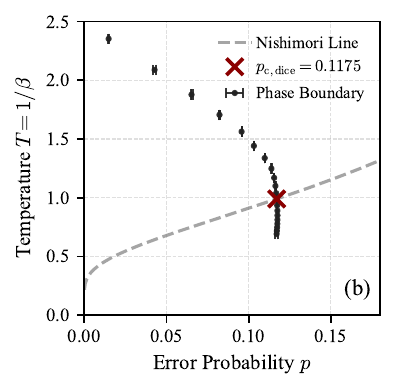}
        \label{fig: phase line for Dice}
    \end{subfigure}
    \caption{Phase diagrams for the dual pair of \textbf{(a)} kagome and \textbf{(b)} dice lattices, showing the multicritical intersections with the Nishimori line (see \cref{tab:appendix_detailed_results}).}
    \label{fig:Phase lines kagome dice}
\end{figure}

\begin{figure}[htbp]
    \centering
    \begin{subfigure}[b]{0.48\textwidth}
        \centering
        \includegraphics[]{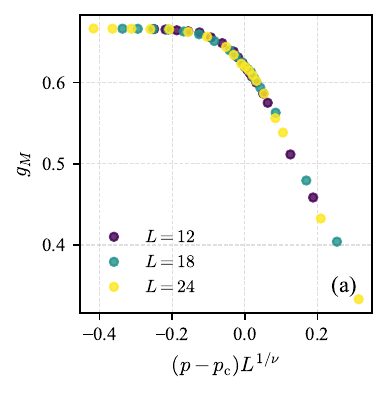}
        \label{fig:data collapse kagome lattice}
    \end{subfigure}
    \hfill 
    \begin{subfigure}[b]{0.48\textwidth}
        \centering
        \includegraphics[]{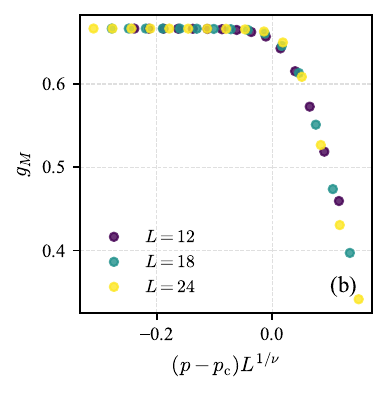}
    \end{subfigure}
    \caption{Data collapse at the Nishimori point for \textbf{(a)} the kagome and \textbf{(b)} the dice lattice (parameters in \cref{tab:appendix_detailed_results}).}
    \label{fig:Data collapse of kagome and dice lattice}
\end{figure}

In this appendix, we compile the phase diagrams, and data collapses for all investigated two-dimensional lattice geometries. Table~\ref{tab:appendix_detailed_results} summarizes the multicritical temperatures, error thresholds, and critical exponents at the Nishimori point obtained via Binder cumulant intersections and data collapse optimizations. Figures~\ref{fig:Phase lines honeycomb triangular}--\ref{fig:Data collapse of kagome and dice lattice} display the corresponding phase boundaries and data collapse curves for the dual lattice pairs.
The obtained critical exponents $\nu$ are consistent with the expected value $1.5$ \cite{merz2002twodimensional,dequeiroz2009location}.

\begin{table}[htbp]
    \centering
    \caption{Summary of numerical parameters extracted from Binder cumulant intersections and data collapse at the Nishimori point across all studied lattice geometries.}
    \label{tab:appendix_detailed_results}
    \begin{tabular}{lcccc}
        \toprule
        \textbf{Lattice} & \textbf{Intersec.} $T_\mathrm{c}$ & \textbf{Intersec.} $p_\mathrm{c}$ & \textbf{Collapse} $p_\mathrm{c}$ & $\nu$ \\
        \midrule
        Square      & $0.9532$ & $0.1093 \pm 0.0006$  & $0.10924 \pm 0.00027$ & $1.47 \pm 0.17$  \\
        Honeycomb   & $0.7497$ & $0.0655 \pm 0.0013$  & $0.0654 \pm 0.0007$   & $1.357 \pm 0.136$ \\
        Triangular  & $1.2236$ & $0.1632 \pm 0.0004$  & $0.1633 \pm 0.0003$   & $1.36 \pm 0.14$  \\
        Kagome      & $0.9101$ & $0.1000 \pm 0.0003$  & $0.1000 \pm 0.0005$   & $1.36 \pm 0.28$  \\
        Dice        & $0.9896$ & $0.11700 \pm 0.00003$ & $0.1175 \pm 0.0006$   & $1.52 \pm 0.10$  \\
        \bottomrule
    \end{tabular}
\end{table}

\newpage
\section{Taylor Expansion of the Dual Threshold Splitting}\label{app:Taylorexpansion}
To analytically investigate the threshold splitting on dual lattices, we start from the Ohzeki duality relation for two-dimensional Calderbank-Shor-Steane surface codes on the Nishimori line \cite{ohzeki2008duality}, defined by
\begin{equation}\label{eq:ohzeki_duality}
    H_2(p_{\mathrm{p}}) + H_2(p_{\mathrm{d}}) - 1 = 0\ ,
\end{equation}
where $p_{\mathrm{p}}$ and $p_{\mathrm{d}}$ denote the error thresholds on the primal and dual lattices, respectively, and $H_2(p) = -p \log_2(p) - (1-p)\log_2(1-p)$ is the binary Shannon entropy.

Using this, the dual threshold can be expressed as an implicit function of the primal threshold $p_{\mathrm{d}}(p_{\mathrm{p}})$ using a Taylor expansion around the self-dual fixed point of the square lattice, $p_{\mathrm{sq}} \approx 0.10924$. On this point, $p_{\mathrm{p}} = p_{\mathrm{d}} = p_{\mathrm{sq}}$ and $H_2(p_{\mathrm{sq}}) = 1/2$. 

The Taylor series expansion up to third order reads
\begin{equation}\label{eq:taylor_expansion_general}
    p_{\mathrm{d}}(p_{\mathrm{sq}} + \Delta p_{\mathrm{p}}) = p_{\mathrm{sq}} + \left. \frac{\mathrm{d}p_{\mathrm{d}}}{\mathrm{d}p_{\mathrm{p}}} \right|_{p_{\mathrm{sq}}} \Delta p_{\mathrm{p}} + \frac{1}{2} \left. \frac{\mathrm{d}^2 p_{\mathrm{d}}}{\mathrm{d}p_{\mathrm{p}}^2} \right|_{p_{\mathrm{sq}}} (\Delta p_{\mathrm{p}})^2 + \frac{1}{6} \left. \frac{\mathrm{d}^3 p_{\mathrm{d}}}{\mathrm{d}p_{\mathrm{p}}^3} \right|_{p_{\mathrm{sq}}} (\Delta p_{\mathrm{p}})^3 + \mathcal{O}((\Delta p_{\mathrm{p}})^4)\ ,
\end{equation}
where $\Delta p_{\mathrm{p}} = p_{\mathrm{p}} - p_{\mathrm{sq}}$ represents the threshold offset of the primal lattice relative to the square lattice.

Differentiating \eqref{eq:ohzeki_duality} with respect to $p_{\mathrm{p}}$ yields
\begin{equation}\label{eq:first_derivative_implicit}
    H_2'(p_{\mathrm{p}}) + H_2'(p_{\mathrm{d}}) \frac{\mathrm{d}p_{\mathrm{d}}}{\mathrm{d}p_{\mathrm{p}}} = 0 \quad \implies \quad \frac{\mathrm{d}p_{\mathrm{d}}}{\mathrm{d}p_{\mathrm{p}}} = -\frac{H_2'(p_{\mathrm{p}})}{H_2'(p_{\mathrm{d}})}\ .
\end{equation}
Evaluating this first derivative at the point $p_{\mathrm{p}} = p_{\mathrm{d}} = p_{\mathrm{sq}}$ gives
\begin{equation}\label{eq:first_derivative_eval}
    \left. \frac{\mathrm{d}p_{\mathrm{d}}}{\mathrm{d}p_{\mathrm{p}}} \right|_{p_{\mathrm{sq}}} = -1\ .
\end{equation}
In first order, the threshold splitting is symmetric, $p_{\mathrm{d}}(p_{\mathrm{sq}} + \Delta p_{\mathrm{p}}) \approx p_{\mathrm{sq}} - \Delta p_{\mathrm{p}}$.\par
To determine the local curvature responsible for the net threshold gain, we differentiate a second time with respect to $p_{\mathrm{p}}$
\begin{equation}\label{eq:second_derivative_implicit}
    H_2''(p_{\mathrm{p}}) + H_2''(p_{\mathrm{d}}) \left( \frac{\mathrm{d}p_{\mathrm{d}}}{\mathrm{d}p_{\mathrm{p}}} \right)^2 + H_2'(p_{\mathrm{d}}) \frac{\mathrm{d}^2 p_{\mathrm{d}}}{\mathrm{d}p_{\mathrm{p}}^2} = 0\ .
\end{equation}
Evaluating at $p_{\mathrm{sq}}$ with $\left( \mathrm{d}p_{\mathrm{d}}/\mathrm{d}p_{\mathrm{p}} \right)^2 = (-1)^2 = 1$ leads to
\begin{equation}\label{eq:second_derivative_eval}
    2 H_2''(p_{\mathrm{sq}}) + H_2'(p_{\mathrm{sq}}) \left. \frac{\mathrm{d}^2 p_{\mathrm{d}}}{\mathrm{d}p_{\mathrm{p}}^2} \right|_{p_{\mathrm{sq}}} = 0 \quad \implies \quad \left. \frac{\mathrm{d}^2 p_{\mathrm{d}}}{\mathrm{d}p_{\mathrm{p}}^2} \right|_{p_{\mathrm{sq}}} = -\frac{2 H_2''(p_{\mathrm{sq}})}{H_2'(p_{\mathrm{sq}})}\ .
\end{equation}
We can continue this procedure equivalently to find the third-order derivative with respect to $p_{\mathrm{p}}$\footnote{This procedure can also be repeated arbitrarily often to find higher orders.}
\begin{equation}\label{eq:third_derivative_implicit}
    H_2'''(p_{\mathrm{p}}) + H_2'''(p_{\mathrm{d}}) \left( \frac{\mathrm{d}p_{\mathrm{d}}}{\mathrm{d}p_{\mathrm{p}}} \right)^3 + 3 H_2''(p_{\mathrm{d}}) \left( \frac{\mathrm{d}p_{\mathrm{d}}}{\mathrm{d}p_{\mathrm{p}}} \right) \left( \frac{\mathrm{d}^2 p_{\mathrm{d}}}{\mathrm{d}p_{\mathrm{p}}^2} \right) + H_2'(p_{\mathrm{d}}) \frac{\mathrm{d}^3 p_{\mathrm{d}}}{\mathrm{d}p_{\mathrm{p}}^3} = 0\ .
\end{equation}
At $p_{\mathrm{p}} = p_{\mathrm{d}} = p_{\mathrm{sq}}$, the third derivatives $H_2'''(p_{\mathrm{sq}})$ cancel out since $\left(\mathrm{d}p_{\mathrm{d}}/\mathrm{d}p_{\mathrm{p}}\right)^3 = (-1)^3 = -1$. Substituting the expression for the second derivative, we obtain
\begin{equation}\label{eq:third_derivative_subst}
    3 H_2''(p_{\mathrm{sq}}) (-1) \left( -\frac{2 H_2''(p_{\mathrm{sq}})}{H_2'(p_{\mathrm{sq}})} \right) + H_2'(p_{\mathrm{sq}}) \left. \frac{\mathrm{d}^3 p_{\mathrm{d}}}{\mathrm{d}p_{\mathrm{p}}^3} \right|_{p_{\mathrm{sq}}} = 0\ ,
\end{equation}
which simplifies to
\begin{equation}\label{eq:third_derivative_eval}
    \left. \frac{\mathrm{d}^3 p_{\mathrm{d}}}{\mathrm{d}p_{\mathrm{p}}^3} \right|_{p_{\mathrm{sq}}} = -6 \frac{\left( H_2''(p_{\mathrm{sq}}) \right)^2}{\left( H_2'(p_{\mathrm{sq}}) \right)^2}\ .
\end{equation}

The derivatives of the binary entropy function $H_2(p)$ are given explicitly by
\begin{align}
    H_2'(p) &= \log_2 \left( \frac{1-p}{p} \right)\ , \\
    H_2''(p) &= -\frac{1}{\ln(2) p (1-p)} < 0\ , \\ 
    H_2'''(p) &= -\frac{2p - 1}{\ln(2) p^2 (1-p)^2}\ .
\end{align}
Because $H_2'(p) > 0$ and $H_2''(p) < 0$ for all $p \in (0, 0.5)$, the second derivative of the dual threshold function $\left. \mathrm{d}^2 p_{\mathrm{d}}/\mathrm{d}p_{\mathrm{p}}^2 \right|_{p_{\mathrm{sq}}}$ is strictly positive.\par
Evaluating the derivatives at the square lattice fixed point $p_{\mathrm{sq}} \approx 0.10924$ yields
\begin{align}
    H_2'(p_{\mathrm{sq}}) &\approx 3.02755\ , \\
    H_2''(p_{\mathrm{sq}}) &\approx -14.8260\ .
\end{align}
Inserting these numerical values into the Taylor coefficients gives
\begin{align}
    c_1 &= \left. \frac{\mathrm{d}p_{\mathrm{d}}}{\mathrm{d}p_{\mathrm{p}}} \right|_{p_{\mathrm{sq}}} = -1\ , \\
    c_2 &= \frac{1}{2} \left. \frac{\mathrm{d}^2 p_{\mathrm{d}}}{\mathrm{d}p_{\mathrm{p}}^2} \right|_{p_{\mathrm{sq}}} = -\frac{H_2''(p_{\mathrm{sq}})}{H_2'(p_{\mathrm{sq}})} \approx +4.8971 \approx +4.90\ , \\
    c_3 &= \frac{1}{6} \left. \frac{\mathrm{d}^3 p_{\mathrm{d}}}{\mathrm{d}p_{\mathrm{p}}^3} \right|_{p_{\mathrm{sq}}} = -\frac{\left( H_2''(p_{\mathrm{sq}}) \right)^2}{\left( H_2'(p_{\mathrm{sq}}) \right)^2} \approx -23.9822 \approx -24.01\ .
\end{align}
Thus, the explicit Taylor expansion predicting the dual threshold as a function of the primal deviation $\Delta p_{\mathrm{p}}$ is
\begin{equation}\label{eq:final_taylor_series}
    p_{\mathrm{d}}(\Delta p_{\mathrm{p}}) \approx p_{\mathrm{sq}} - \Delta p_{\mathrm{p}} + 4.90 (\Delta p_{\mathrm{p}})^2 - 24.01 (\Delta p_{\mathrm{p}})^3 + \mathcal{O}((\Delta p_{\mathrm{p}})^4)\ .
\end{equation}
The strictly positive leading-order correction term $+4.90 (\Delta p_{\mathrm{p}})^2$ analytically demonstrates that any asymmetry in lattice connectivity results in a net systematic increase in the sum of thresholds $p_{\mathrm{p}} + p_{\mathrm{d}} > 2 p_{\mathrm{sq}}$.\par

\newpage
\bibliography{bibliographie.bib}
\end{document}